# On the Hall-mediated resistive tearing instability of highly elongated current sheets

Grigory Vekstein[1,2,a]

[1]Jodrell Bank Centre for Astrophysics, School of Physics and Astronomy, University of Manchester, UK

[2] National Aviation Academy, Baku, Republic of Azerbaijan

**Abstract**. The present paper provides comprehensive description of various regimes involved in the two-fluid model of the resistive tearing instability. These include two novel regimes of this instability, which correspond to the long-wave modes that can develop in a highly-elongated current sheet. This issue is relevant to the study of fast magnetic reconnection and magnetic turbulence in magnetohydrodynamic (MHD) objects with a large value of the Lundquist number.

## 1. Introduction

A renewed interest in the properties of a long wave-length tearing modes has been prompted by the observation that the so-called plasmoid instability (the tearing mode developing in evolving current sheets) can play significant role in fast magnetic reconnection[1-4] and magnetic turbulence[5,6] occurring in a highly conducting medium. In this context an important issue is revealing the fastest mode (a mode with the maximal linear instability growth rate), which kick-starts the subsequent process of nonlinear reconnection. It is well-known from the classical Furth-Killeen-Rosenbluth (FKR) theory of the tearing instability[7] that its growth rate increases with the mode wave-length. However, this theory is not applicable to a highly elongated current sheet, which may form in a system with a very large Lundquist number. The respective generalisation of the FKR theory has been developed in Ref.8 (see also Refs.9.10). It should be noted, however, that all the results presented in Refs.7-10 were obtained in the framework of the standard single-fluid MHD. On the other hand, as it was originally pointed out in Refs.11,12, the growth rate of the tearing instability can be significantly increased by the Hall effect, which originates from the two-fluid (electrons and ions) description of the plasma. Then magnetic field is advected by the flow of the light electron plasma component rather than by the bulk flow of matter associated with the much heavier ions (as it is the case in the single-fluid MHD). Although this issue has been already discussed in numerous publications (see, e.g., Ref.13 and the list of references therein), a complete explicit verification of all possible Hall-mediated instability regimes and transitions between them is still lacking. Therefore, the present paper aims to fill this gap by providing a unified description of the two-fluid resistive tearing instability in an inviscid highly conducting plasma. The suggested scheme is valid for arbitrary values of the main relevant parameters, which are the plasma $\beta$, the ion skin depth $d_i = c/\omega_{pi}$ (it signifies the strength of the Hall effect), and a mode wave-vector $k$.

[a]Email: g.vekstein@manchester.ac.uk

A value of the latter determines whether the mode growth can be described by the FKR-type theory (the so-called constant-$\psi$ regime[7]), or it requires the non-constant-$\psi$ approach (the so-called Coppi regime[8]), which is relevant to a very long-wave modes. Thus, this article presents a generalisation of Ref.14, which dealt only with the constant-$\psi$ case.

As an example, here we consider a particular case of the well-known Harris-type force-free magnetic field, though the obtained results can be easily modified for any other initial magnetic equilibrium. Thus, a uniform plasma of density $n_i = n_e = n_0$, and thermal pressure $P = P_0$, is immersed in the magnetic field

$$\vec{B}^{(0)} = \left[0, B_0 th\left(\frac{x}{L}\right), B_0 ch^{-1}\left(\frac{x}{L}\right)\right] \tag{1}$$

For tearing modes with a wave-vector $\vec{k}$ directed along the $y$-axis the system remains invariant along the $z$-axis. Hence, it what follows it is convenient to introduce the flux function $\Psi(x, y)$ as

$$\vec{B}(x, y) = [\vec{\nabla}\Psi(x, y) \times \hat{z}] + B_z(x, y)\vec{z}, \tag{2}$$

so that $B_x = \partial\Psi/\partial y$ and $B_y = -\partial\Psi/\partial x$. Thus, the field (1) corresponds to the flux function

$$\Psi_0(x) = -B_0 L \ln\left[ch\left(\frac{x}{L}\right)\right] \tag{3}$$

The tearing perturbation results in $\Psi(x, y) = \Psi_0(x) + \Psi_1(x, y)$, where for a single Fourier harmonic

$$\Psi_1(x, y) = \psi(x)\cos ky. \tag{4}$$

Following the standard procedure[7], the first step is derivation of the so-called "external" equilibrium solution, which in a low-$\beta$ plasma is a new force-free magnetic configuration. The latter should satisfy the Grad-Shafranov equation

$$B_z(x, y) = F(\Psi), \nabla^2\Psi + \frac{1}{2}\frac{d}{d\Psi}[F^2(\Psi)] = 0. \tag{5}$$

In the case of the Harris equilibrium (1) $F(\Psi) = F_0(\Psi) = B_0 \exp\left(\frac{\Psi}{B_0 L}\right)$. In the linear approximation with respect to $\Psi_1 \ll \Psi_0$ the function $F(\Psi)$ retains its form[15], hence,

$$\frac{d}{d\Psi}[F^2(\Psi)] = \frac{2B_0}{L}\exp\left(\frac{2(\Psi_0 + \Psi_1)}{B_0 L}\right) \approx \frac{2B_0}{L}\exp\left(\frac{2\Psi_0}{B_0 L}\right)\left(1 + \frac{2\Psi_1}{B_0 L}\right) = \frac{2B_0}{Lch^2(x/L)}\left(1 + \frac{2\Psi_1}{B_0 L}\right),$$

so the linearized version of (5) yields the following equation for the function $\psi(x)$:

$$\frac{d^2\psi}{dx^2} - k^2\psi + \frac{2}{L^2 ch^2(x/L)}\psi = 0.$$

Its appropriate solution (the one that tends to zero at infinity) reads[7]

$$\psi(x) = \varepsilon B_0 L \exp(\mp kx)\left[1 \pm \frac{th(x/L)}{kL}\right], \quad \varepsilon \ll 1 \tag{6}$$

It is singular at $x = 0$, which results in the following expression for the tearing instability parameter [7] $\Delta'$:

$$\Delta' = \frac{\psi'(+0) - \psi'(-0)}{\psi(0)} = \frac{2}{L}\left(\frac{1}{kL} - kL\right) \tag{7}$$

Therefore, as seen from (7), the tearing instability ($\Delta' > 0$) corresponds to the long-wave perturbations with $kL < 1$. Note also that perturbation (6) of the flux function $\Psi(x, y)$ yields the respective change in the $z$ - component of the magnetic field:

$$B_z^{(1)} = \frac{\varepsilon B_0}{ch(x/L)} \exp(\mp kx)\left[1 \pm \frac{th(x/L)}{kL}\right] \cos ky \tag{8}$$

The tearing instability growth rate, $\gamma(k)$, is determined by the plasma dynamics inside the current sheet, which forms due to the singularity of solution (6) at $x = 0$. The respective solution, the "internal" one, is governed by the equation of plasma motion:

$$\rho \frac{d\vec{V}}{dt} = \frac{1}{c}(\vec{j} \times \vec{B}) - \vec{\nabla}P, \quad \vec{j} = \frac{c}{4\pi}(\vec{\nabla} \times \vec{B}), \tag{9}$$

and the magnetic induction equation:

$$\frac{\partial \vec{B}}{\partial t} = \vec{\nabla} \times (\vec{V} \times \vec{B}) + \eta \nabla^2 \vec{B} - \frac{1}{ne}\vec{\nabla} \times (\vec{j} \times \vec{B}), \tag{10}$$

where the last term on the r.h.s. of Eq.(10) accounts for the Hall effect. For the sake of simplicity, we assume that the ion plasma component is cold, hence, the effect of the ion gyro-viscosity is not included in Eq.(9). Then, the first task is the linearization of Eqs.(9,10) in respect to small perturbations $\Psi_1, B_z^{(1)}$ and $\vec{V}$ (note that plasma is initially at rest). The latter can be represented as

$$\vec{V}(x, y) = (\vec{\nabla}\Phi(x, y) \times \hat{z}) + \vec{\nabla}X(x, y) + V_z(x, y)\hat{z}, \tag{11}$$

where the stream-function $\Phi$ and the velocity potential $X$ correspond, respectively, to the rotational and compressional components of the plasma flow. Thus, by using the continuity equation, one can express the pressure perturbation in terms of $X$, which in the linear approximation yields

$P_1 = -\frac{\Gamma}{\gamma} P_0 (\vec{\nabla} \cdot \vec{V}) = -\frac{\Gamma}{\gamma} P_0 \nabla^2 X$, where $\Gamma$ is the adiabatic index. By defining the plasma beta as

$\beta \equiv \frac{4\pi \Gamma P_0}{B_0^2}$, this pressure perturbation can be re-written as $P_1 = -\beta \frac{B_0^2}{4\pi\gamma} \nabla^2 X$.

Furthermore, since the width of the central current sheet is much smaller than the characteristic length scale $L$, in the internal solution one can put $\Psi_0(x) \approx -B_0 x^2 / 2L, B_z^{(0)}(x) \approx B_0$. After that Eq.(9) yields the following set of equations for $\Phi, X$ and $V_z$:

$$\gamma \nabla^2 \Phi = \frac{B_0}{4\pi\rho} \left( \frac{x}{L} \frac{\partial}{\partial y} \nabla^2 \Psi_1 \right), \tag{12}$$

$$\gamma \nabla^2 X = -\frac{B_0}{4\pi\rho} \left[ \frac{\beta B_0}{\gamma} \nabla^2 (\nabla^2 X) + \nabla^2 \left( B_z^{(1)} - \frac{\Psi_1}{L} \right) \right], \tag{13}$$

$$\gamma V_z = \frac{B_0}{4\pi\rho} \cdot \frac{x}{L} \frac{\partial}{\partial y} \left( B_z^{(1)} - \frac{\Psi_1}{L} \right) \tag{14}$$

Similarly, the linearized magnetic induction equation (10) results in two equations for $\Psi_1$ and $B_z^{(1)}$:

$$\gamma \Psi_1 = V_x B_0 \frac{x}{L} + \eta \nabla^2 \Psi_1 + \frac{cB_0}{4\pi n e} \frac{x}{L} \frac{\partial}{\partial y} \left( B_z^{(1)} - \frac{\Psi_1}{L} \right), \tag{15}$$

$$\gamma B_z^{(1)} = V_x B_0 \frac{x}{L^2} - B_0 \nabla^2 \chi + B_0 \frac{x}{L} \frac{\partial V_z}{\partial y} + \eta \nabla^2 B_z^{(1)} - \frac{cB_0}{4\pi n e} \frac{x}{L} \frac{\partial}{\partial y} \nabla^2 \Psi_1 \tag{16}$$

As seen from the last term in Eq.(15), the Hall effect is associated with the magnetic field component $B_z^{(H)} = B_z^{(1)} - \Psi_1 / L$. Thus, under the flux function perturbation $\Psi_1$ of the form (4), with $\psi(x)$ being an even function of $x$, it follows from (15) that the Hall magnetic field takes the form

$$B_z^{(H)} = b(x) \sin ky, \tag{17}$$

with an odd $b(x)$. Thus, this a quadrupole magnetic field, which is generated by the last, Hall term, in Eq.(16). The symmetry of all other functions becomes then transparent from Eqs.(12-14):

$$\Phi(x, y) = \varphi(x) \sin ky, X(x, y) = \chi(x) \sin ky, V_z(x, y) = v(x) \cos ky,$$

where $\varphi, \chi$ are odd, and $v$ is an even functions of $x$.

After introducing non-dimensional variables and functions by the re-scaling:

$x \to x/L, k \to kL, \Psi_1 \to \Psi_1/B_0 L, B_z^{(H)} \to B_z^{(H)}/B_0, \Phi \to \Phi/\gamma L^2,$
$X \to X/\gamma L^2, V_z \to V_z/\gamma L,$

one can reduce Eqs.(12-16) to a simplified set of ordinary differential equations for functions $\psi(x), b(x), \varphi(x), \chi(x)$ as

$$\psi = kx\varphi + \frac{1}{(\gamma\tau_A)S}\psi'' + \frac{dkx}{(\gamma\tau_A)}b, \tag{18}$$

$$b = -\chi'' - \frac{k^2 x^2}{(\gamma\tau_A)^2}b + \frac{1}{(\gamma\tau_A)S}b'' + \frac{dkx}{(\gamma\tau_A)}\psi'', \tag{19}$$

$$\varphi'' = -\frac{kx}{(\gamma\tau_A)^2}\psi'', \tag{20}$$

$$\chi'' = -\frac{1}{(\gamma\tau_A)^2}[\beta\chi^{(IV)} - b''] \tag{21}$$

Here $\tau_A = L/V_A$ is the global Alfven transit time defined by the Alfven velocity $V_A = B_0/\sqrt{4\pi\rho}$, $S = LV_A/\eta$ is the Lundquist number (which is assumed to be large, $S \gg 1$), and $d \equiv d_i/L$ is the scaled ion skin depth (for most of applications it is small, $d < 1$). Since under a large value of the Lundquist number the width, $(\Delta x)$, of the internal resistive current is small, $(\Delta x) \ll 1$ (see below), and for the unstable modes $k < 1$ [see Eq.(7)], in the derivation of the above equations the approximation $\nabla^2 = \frac{d^2}{d^2 x} - k^2 \approx \frac{d^2}{d^2 x}$ is used. Note also that only the rotational flow component is significant in the magnetic field advection by the bulk plasma flow [the first term on the r.h.s. of Eq.(18)].

In what follows we first briefly re-consider the FKR-type (the constant-psi) regimes of the tearing instability (Section 2), leaving the Coppi-type (the non-constant-psi) regimes for Section 3. Implications of these results on the issue of the Hall-mediated plasmoid instability is discussed in Section 4.

## 2. The FKR-type (the constant-psi) regimes.

In this case, without loss of generality, one can put $\psi = 1$, so the matching condition of the internal and external solutions then reads $\Delta' = \int \psi'' dx$. For the current sheet of width $(\Delta x)$ this integral can be estimated as $\psi'' \cdot (\Delta x)$, which yields

$$\psi'' \sim \Delta' \cdot (\Delta x)^{-1} \sim k^{-1}(\Delta x)^{-1} \tag{22}$$

Further on, by using the equation of motion (20), one can estimate the stream function in the internal solution. Thus, $\varphi'' \sim \varphi \cdot (\Delta x)^{-2}$, while, according to (22), $x\psi'' \sim (\Delta x) \cdot \psi'' \sim \Delta' \sim k^{-1}$. Hence, it follows from (20) that

$$\varphi \sim (\gamma \tau_A)^{-2} \cdot (\Delta x)^2 \qquad (23)$$

Consider now the magnetic induction equation (18), assuming first that the Hall parameter $d$ is small enough (see below for the exact criterion) to make the last, Hall term, in (18) insignificant. Then, the ongoing magnetic reconnection, the pace of which is defined by the l.h.s. of this equation, is supported by both the plasma resistivity [the second term on the r.h.s. of (18)] and the magnetic field advection into the current sheet by the bulk plasma flow [the first term on the r.h.s. of (18)]. Therefore, all relevant three terms in Eq.(18) should be of the same order of magnitude. Thus, by comparing the first two with help of (22) and (23), one gets

$$kx\varphi \sim k(\Delta x)(\gamma \tau_A)^{-2}(\Delta x)^2 \sim 1 \Rightarrow (\Delta x) \sim (\gamma \tau_A)^{2/3} k^{-1/3}, \qquad (24)$$

and equating the first and the third terms yields

$$(\gamma \tau_A)^{-1} S^{-1} \psi'' \sim (\gamma \tau_A)^{-1} S^{-1} k^{-1} (\Delta x)^{-1} \sim 1 \qquad (25)$$

It follows then from (24-25) that

$$(\gamma \tau_A) \sim S^{-3/5} k^{-2/5}, (\Delta x) \sim S^{-2/5} k^{-3/5}, \qquad (26)$$

re-producing the well-known scaling of the FKR theory[7].

Consider now what restrictions apply to this solution by the imposed "constant-psi" assumption. Clearly, the variation of the flux function across the current sheet, which can be estimated as $\Delta \psi \sim \psi'' \cdot (\Delta x)^2 \sim \Delta' \cdot (\Delta x)$, should remain small, i.e. $\Delta' \cdot (\Delta x) < 1$. For $\Delta' \sim k^{-1}$ and $(\Delta x)$ given in Eq.(26), this requirement takes the form

$$k > k_* \sim S^{-1/4} \qquad (27)$$

The case of the long-wave modes with $k < k_*$ is discussed below in Section 3.

Consider now what value of the parameter $d$ is required to make the Hall effect coming into play. Such a situation occurs when the Hall term in Eq.(18), which describes additional advection of the magnetic field into the current sheet by the flow of electrons caused by the quadrupole magnetic field perturbation $b$, becomes comparable with the first term on the r.h.s. of Eq.(18), which is due to the standard magnetic field advection by the bulk plasma flow. This quadrupole field is, in its own turn, generated by the Hall term in Eq.(19), and the resulting magnitude of $b$ is determined by the balance between this source term and the other terms on the r.h.s. of (19). They are due, respectively, to the plasma compression, the bulk velocity component $V_z$, and the resistive field diffusion. Therefore, in order to evaluate their relative role, it is necessary to invoke Eq.(21), which describes the compressional part of the plasma flow. The latter is driven by the magnetic force due to the quadrupole field $b$, which is balanced either by the thermal pressure force (the first term on the r.h.s. of this equation) or by the plasma inertia [the l.h.s. of Eq.(21)]. A simple comparison of the two reveals that the inertial effect is significant only under extremely small values of the plasma beta [$\beta < (\Delta x)^2 (\gamma \tau_A)^2$], in which case the respective Hall-mediated regime of instability takes place only under unrealistically large values of the Hall parameter $d$ [14]. Therefore, in what follows it is assumed that the magnetic force under consideration is balanced by the thermal pressure, so, according to (21),

$$\chi'' = \frac{b}{\beta} \qquad (28)$$

Hence, if the plasma beta is not too high (see below), the plasma compression is the major factor that determines magnitude of the quadrupole field. Then, a comparison of the first and last terms on the r.h.s. of Eq.(21) [with help of (22) and (26)] yields

$$b \sim \beta (\gamma \tau_A)^{-1} dk(\Delta x)\psi''' \sim \beta d S^{3/5} k^{2/5} \qquad (29)$$

Under this quadrupole field the Hall term in Eq.(18) can be estimate as $(\gamma \tau_A)^{-1} dkxb \sim \beta d^2 k^{6/5} S^{4/5}$, which becomes comparable to other terms in (18) at

$$d \sim d_1 \sim \beta^{-1/2} S^{-2/5} k^{-3/5} \qquad (30)$$

Therefore, under $d > d_1$, all properties of the tearing instability are determined by the Hall effect. In this case Eq.(22) still holds, but, instead of Eq.(24), we now have from (18), (19) and (28) that

$$(\gamma \tau_A)^{-1} d(\Delta x)b \sim 1, \qquad (31)$$

and

$$b \sim \beta (\gamma \tau_A)^{-1} dk(\Delta x)\psi''' \sim \beta (\gamma \tau_A)^{-1} d \qquad (32)$$

Altogether, (25) and (31-32) yield

$$(\gamma \tau_A) \sim \beta^{1/3} d^{2/3} S^{-1/3}, (\Delta x) \sim \beta^{-1/3} d^{-2/3} S^{-2/3} k^{-1} \qquad (33)$$

This regime, however, cannot survive under increasing plasma beta. Indeed, according to Eq.(28), the resulting reduction in the plasma compressibility makes its role in Eq.(19) insignificant (plasma becomes effectively incompressible), and the magnitude of the quadrupole magnetic field is then determined by its resistive diffusion. Such a transition occurs when the compressional and resistive terms in Eq.(19) become roughly equal. Hence, $\chi'' = \beta^{-1} b \sim (\gamma \tau_A)^{-1} S^{-1} b'' \sim (\gamma \tau_A)^{-1} S^{-1} (\Delta x)^{-2} b$, and with help of (33) it yields

$$\beta \sim S^{-1/2} d^{-1/2} k^{-3/2} \Rightarrow d \sim d_2 \sim S^{-1} \beta^{-2} k^{-3} \qquad (34)$$

(note that $d_2 > d_1$ if $k > k_*$ and $\beta \leq 1$).

Thus, in order to obtain the instability regime under $d > d_2$, one should supplement Eqs. (22), (25) and (31) with the relation accounting for the balance of the resistive and Hall terms in Eq.(19). This relation reads

$$(\gamma\tau_A)^{-1} S^{-1} (\Delta x)^{-2} b \sim (\gamma\tau_A)^{-1} dk(\Delta x)\psi'' \sim (\gamma\tau_A)^{-1} d \Rightarrow b \sim Sd(\Delta x)^2, \quad (35)$$

so finally one gets for this regime that

$$(\gamma\tau_A) \sim S^{-1/2} d^{1/2} k^{-1/2}, (\Delta x) \sim S^{-1/2} d^{-1/2} k^{-1/2} \quad (36)$$

This is the well-known scaling originally derived[16] in the framework of the Electron MHD (EMHD), when the ion component is assumed to be immobile from the outset. In terms of Eqs.(20-21) EMHD corresponds to the limit $d \to \infty$. However, this instability regime can be achieved even under a small value of the Hall parameter, $d \ll 1$ (see Eq.(34) and Fig.1), when the main body of a system is still governed by the single-fluid MHD equations. Indeed, in this case the internal solution acquires a double structure, when the Hall effect remains significant only inside the region $x \leq x_H$ [$(\Delta x) \ll x_H \ll 1$], the size of which can be estimated in the following way. At the distances $(\Delta x) < x < x_H$ the plasma resistivity plays no role, but advection of magnetic field in Eq.(18) is still due to the Hall term, hence

$$(\gamma\tau_A)^{-1} dkxb \sim 1 \Rightarrow b \sim (\gamma\tau_A)(kdx)^{-1} \quad (37)$$

This level of the quadrupole field should be provided by the Hall term in Eq.(19), where the necessary balance is now maintained by the second term on the r.h.s. of this equation. The latter originates from the $V_z$ component of the bulk plasma flow [see Eqs.(14) and (16)]. Thus, with help of (37), one gets

$$(\gamma\tau_A)^{-2} (kx)^2 b \sim (\gamma\tau_A)^{-1} dkx\psi'' \Rightarrow \psi'' \sim d^{-2} \quad (38)$$

On the other hand, the very same electric current (38) drives the bulk plasma flow, hence, according to (20), (36), and (38), $\varphi'' \sim (\gamma\tau_A)^{-2} kx\psi'' \Rightarrow \varphi \sim Sd^{-3} k^2 x^3$. Thus, the magnetic field advection by the bulk plasma flow [the first term on the r.h.s. of Eq.(18)] grows with distance $x$ as $kx\varphi \sim Sd^{-3} k^3 x^4$. Therefore, the bulk flow advection takes over from the Hall one where $kx\varphi \sim 1$, which occurs at

$$x \sim x_H \sim S^{-1/4} d^{3/4} k^{-3/4} \quad (39)$$

As seen from (36) and (39), the assumed inequality $x_H > (\Delta x)$ holds if $d > S^{-1/5} k^{1/5}$, which is the case for the Hall-mediated regime of instability (see Fig.1). At the same time $x_H < 1$ if the Hall parameter $d \leq 1$, and the constant-psi requirement $k > k_* \sim S^{-1/4}$ holds.

Note also that this ideal Hall-MHD sublayer does not affect the matching condition of the internal and external solutions. Indeed, the respective integral [see Eqs.(38-39)] is
$\int \psi'' dx \sim \psi'' \cdot x_H \sim S^{-1/4} d^{-5/4} k^{-3/4} << \Delta' \sim k^{-1}$ if $d > S^{-1/5} k^{1/5}$.

In summary, all three constant-psi Hall-MHD regimes of the tearing instability are depicted in Fig.1. The regime 1 corresponds to small values of the Hall parameter, which makes this regime identical to the single-fluid regime of the FKR theory[7] [see Eq.(26)]. The regime 2(a) [see Eq.(36)] reproduces the scaling of the Electron MHD tearing mode[16]. Finally, the regime 2(b) [see Eq.(33)] is of an intermediate nature. In this case advection of the magnetic field is caused by the Hall effect, but the strength of the respective quadrupole magnetic field is determined by the plasma compressibility.

It's worth noting that since the Hall effect brings about the additional advection of the magnetic field towards the reconnection site, it results in the reduction of the width ($\Delta x$) of the resistive current sheet. This is favourable for the applicability of the constant-psi approximation. Therefore, in the diagram of Fig.1, which is drawn for the case of $k > k_* \sim S^{-1/4}$, all three regimes are well in the realm of the constant-psi simplification.

## 3. The Coppi-type regimes (the long wave-length tearing modes)

For a long-wave mode with $k < k_*$ the constant-psi approximation becomes non-applicable. Therefore, in this case one has to distinguish between the $\psi_e(0)$, which is the limit of the external solution at $x \to 0$, and $\psi_i(0)$, which is defined by the internal solution. Note that the former determines the tearing parameter $\Delta'$, while the latter defines the amount of reconnected magnetic flux confined in magnetic islands formed inside the current sheet.

In what follows, we put, as before, $\psi_e(0) = 1$, and denote $\psi_i(0)$ as $\psi_i$. Assuming that the Hall parameter $d$ is sufficiently small (see below) we consider first the single-fluid regime of the tearing instability (the regime 1). In this case the above-derived relations (22-23) remain unchanged, while in Eq.(18) the reconnected flux $\psi$, which was equal to 1 in the constant-psi approximation, is now equal to $\psi_i >> 1$ (see below). Hence, instead of (24), one gets

$$kx\varphi \sim k(\Delta x)(\gamma \tau_A)^{-2}(\Delta x)^2 \sim \psi_i \Rightarrow (\Delta x) \sim (\gamma \tau_A)^{2/3} k^{-1/3} \psi_i^{1/3} \qquad (40)$$

Equating [with help of (22)] the first and the third terms of Eq.(18) yields

$$\psi_i \sim (\gamma \tau_A)^{-1} S^{-1} \psi'' \sim (\gamma \tau_A)^{-1} S^{-1} k^{-1} (\Delta x)^{-1} \qquad (41)$$

Furthermore, since inside the current sheet the flux function is not a constant, its total variation across the current sheet, which is equal to $\Delta \psi = \psi_i - \psi_e = \psi_i - 1 \approx \psi_i$, should be accounted for as

$$\Delta \psi \approx \psi_i \sim \psi'' \cdot (\Delta x)^2 \sim (\Delta x) k^{-1} \qquad (42)$$

It follows then from (40-42) that

$$(\gamma\tau_A) \sim S^{-1/3}k^{2/3}, (\Delta x) \sim S^{-1/3}k^{-1/3}, \psi_i \sim S^{-1/3}k^{-4/3} \, , \qquad (43)$$

which reproduces the results originally obtained in Ref.8. It turns out that the scaling (43) has a universal character[10], which does not depend on the particular form of the tearing parameter $\Delta'(k)$ (providing that the latter is large enough to bring the mode into the non-constant-psi regime: $\Delta' \cdot (\Delta x) > 1$). An obvious requirement, $(\Delta x) < 1$, yields the following validity range for this regime:

$$S^{-1} < k < k_* \sim S^{-1/4}, \qquad (44)$$

so this condition is assumed to hold in what follows.

As seen from (43) and (44), the reconnected magnetic flux is large indeed: $\psi_i \gg 1$. Note also that the restriction (44) is necessary for justifying the quasi-static assumption imposed on the external solution. The point is that the characteristic spatial scale for a mode with a wave-number $k \ll 1$ is equal to $\lambda \sim k^{-1}L \gg L$. Therefore, the respective inertial time-scale is $\tau_A^{(\lambda)} \sim k^{-1}\tau_A \gg \tau_A$, which makes the quasi-static criterion more restrictive, namely

$$\gamma(k)\tau_A^{(\lambda)} \sim \gamma(k)\tau_A k^{-1} < 1. \qquad (45)$$

As seen from (43), it is satisfied under the condition (44).

This solution can be used as a starting point for the Hall-mediated regimes. The first task, similar to that in Section 2, is to find out, by using now scaling (43), the threshold value of the Hall parameter $d$. Thus, if the plasma beta is not too small (as shown below, it requires $\beta \geq 1$), the plasma can be assumed incompressible. Then, in Eq.(19) for $b$, the generating Hall term is balanced by the resistive one, so the established magnitude of the quadrupole field is given in Eq.(35). By inserting it into the Hall term in Eq.(18), one gets

$$\frac{dkx}{(\gamma\tau_A)}b \sim \frac{d^2 Sk(\Delta x)^3}{(\gamma\tau_A)} \qquad (46)$$

Under the scaling (43) it amounts to $d^2 S^{1/3}k^{-2/3}$, and it should be compared with the l.h.s. of Eq.(18), which is the reconnected magnetic flux $\psi_i$ given in (43). Therefore, the sought after transition to the Hall-mediated regime [the regime 2(a)] takes place at

$$d = d_1 \sim S^{-1/3}k^{-1/3} < 1. \qquad (47)$$

Hence, at $d > d_1$, in the Hall-MHD regime 2(a) of the instability, it follows from Eqs.(18) and (46) that $\psi_i \sim (\gamma\tau_A)^{-1}d^2 Sk(\Delta x)^3$, which together with Eqs. (41-42) yields

$$(\gamma\tau_A) \sim kd, (\Delta x) \sim S^{-1/2}(kd)^{-1/2}, \psi_i \sim S^{-1/2}d^{-1/2}k^{-3/2} \qquad (48)$$

Quite remarkably, in this regime the instability growth rate does not depend et al on a magnitude of the Lundquist number, though the essentially involved magnetic reconnection is not possible without a finite plasma resistivity. As seen from (48), the quasi-static requirement (45) is satisfied if $d<1$. Otherwise, the plasma inertia affects the external solution[17], reducing the tearing parameter $\Delta'$.

Similarly to the constant-psi case, this Hall-mediated regime is also associated with a double structure of the internal solution. Thus, instead of (37), we now get from Eq.(18) that

$$(\gamma\tau_A)^{-1}dkxb \sim \psi_i \Rightarrow b \sim (\gamma\tau_A)\psi_i(dkx)^{-1},$$

which modifies Eq.(38) into

$$\psi'' \sim d^{-2}\psi_i \sim S^{-1/2}d^{-5/2}k^{-3/2} \qquad (49)$$

It follows then from Eqs.(20) and (48-49) that

$$\varphi'' \sim (\gamma\tau_A)^{-2}kx\psi'' \Rightarrow \varphi \sim S^{-1/2}d^{-9/2}k^{-5/2}x^3,$$

so in Eq.(18) the magnetic field influx associated with the bulk plasma flow, $kx\varphi$, becomes significant, i.e. comparable to $\psi_i$ [see Eq.(48)] at

$$x \sim x_H \sim d \qquad (50)$$

Therefore, in this regime the width of the ideal sublayer is sufficiently small, $x_H<1$, while, on the other hand, it does exceed the width $(\Delta x)$ of the resistive sublayer. Indeed, according to (48), $(\Delta x)/x_H \sim S^{-1/2}d^{-3/2}k^{-1/2} < 1$ if $d > d_1$ [see Eq.(47)].

Note that, as before, the ideal Hall-MHD sublayer does not affect the matching condition defined by the tearing parameter $\Delta'$. Indeed, across this sublayer
$$\int \psi'' dx \sim \psi'' \cdot x_H \sim S^{-1/2}d^{-3/2}k^{-3/2} \ll \Delta' \sim k^{-1}.$$

As mentioned in Section 2, inclusion of the Hall effect is favourable for the constant-psi condition. Thus, further increase of the Hall parameter d made the width $(\Delta x)$ of resistive sublayer [see Eq.(48)] small enough to recover validity of the constant-psi regime. Indeed, as follows from Eq.(48), at

$$d \sim d_2 \sim S^{-1}k^{-3}, \qquad (51)$$

(note that $d_2 > d_1$ when $k < k_* \sim S^{-1/4}$) the magnitude of $\psi_i$ becomes of order of unity, which indicates the transition to the constant-psi regime [the regime 3(a)], which is identical to the EMHD regime 2(a) discussed in Section 2. Hence, at $d > d_2$ the scaling (36) holds, so:

$$(\gamma\tau_A) \sim S^{-1/2}d^{1/2}k^{-1/2}, (\Delta x) \sim S^{-1/2}d^{-1/2}k^{-1/2} \qquad (52)$$

Clearly, as all regimes under consideration involve the external plasma equilibrium (6), the quasi-static condition (45) should be satisfied, which in the case of (52) requires $d < d_{max} \sim Sk^3$. Therefore, the regime 3(a) does exist if $d_2 < d_{max}$, i.e.

$$S^{-1/3} < k < k_* \sim S^{-1/4} , \qquad (53)$$

( the case of $k < S^{-1/3}$ is discussed below).

It is worth mentioning that the regime 2(a), similarly to the regime 3(a), does not involve the bulk plasma motion. Therefore, the instability scalings for these regimes, (48) and (52) respectively, can be re-formulated in terms of the Electron MHD. Thus, since in the EMHD the plasma dynamics is associated with the whistler mode[16], the characteristic dynamical time-scale becomes equal to $\tau_w = L^2 / V_A d_i = \tau_A d^{-1}$. This yields the respective "whistler" Lundquist number $S_w \equiv \tau_\eta / \tau_w = Sd$. Hence, instead of (48), one now gets for the regime 2(a):

$$(\gamma \tau_w) \sim k, (\Delta x) \sim S_w^{-1/2} k^{-1/2}, \psi_i \sim S_w^{-1/2} k^{-3/2}$$

Similarly, for the regime 3(a) the scaling (52) transforms into

$$(\gamma \tau_w) \sim S_w^{-1/2} k^{-1/2}, (\Delta x) \sim S_w^{-1/2} k^{-1/2}$$

Furthermore, the regime 2(a) can be termed as the non-constant-psi EMHD regime, while the regime 3(a) as the constant-psi EMH D regime. The boundary between the two, which is given by Eq.(51), takes the form $k \sim S_w^{-1/3}$, so the regime 2(a) corresponds to $k < S_w^{-1/3}$, and the regime 3(a) to $k > S_w^{-1/3}$. Note, however, that in the framework of the EMHD, when the Hall parameter $d \to \infty$, this boundary is more of a formal character. The point is that the quasi-static nature of the external solution becomes violated well before the constant-psi boundary is reached. Indeed, since the whistler time-scale $\tau_w \propto L^2$, the respective quasi-static condition, analogous to (45) in the case of the MHD, reads $\gamma \tilde{\tau}_w < 1$, with $\tilde{\tau}_w \sim \tau_w k^{-2}$. Hence, for $\gamma \sim \tau_w^{-1} S_w^{-1/2} k^{-1/2}$, it yields $k > S_w^{-1/5}$, which is much larger than $S_w^{-1/3}$ when $S_w \gg 1$. Therefore, as pointed out in Ref.17, in derivation of the instability growth rate for modes with $k < S_w^{-1/5}$, one should take into account the reduction of the tearing parameter $\Delta'$.

Consider now the effect of the plasma compressibility, which becomes significant at $\beta < 1$. First, note that in the constant-psi domain one can still use here the results of Section 2. Thus, as seen in Fig.1, at $\beta \sim (S^{-1} k^{-3} d^{-1})^{1/2}$ the transition to the intermediate regime 2(b) of Section2 takes place. In the notations of Section 3 (see Fig.2 ) it corresponds to $\beta \sim \tilde{\beta} \sim (d_2 / d)^{1/2}$ [see Eq.(51)] , hence, for $\beta < \tilde{\beta}$ the regime, defined in Fig.2 as the regime 3(b), follows the scaling (33):

$$(\gamma \tau_A) \sim \beta^{1/3} d^{2/3} S^{-1/3}, (\Delta x) \sim \beta^{-1/3} d^{-2/3} S^{-2/3} k^{-1} \qquad (54)$$

However, further reduction in $\beta$ makes the effect of the plasma compressibility stronger, which results in the increased width $(\Delta x)$ of the resistive layer. Therefore, eventually, the limit of the constant-psi condition, which reads $(\Delta x) k^{-1} \sim 1$, is reached, according to (54), at

$$d \sim \tilde{d}_2 \sim \beta^{-1/2} S^{-1} k^{-3} \tag{55}$$

Therefore, the domain $d < \tilde{d}_2$ in Fig.2 corresponds to the intermediate non-constant-psi Hall-MHD regime [regime 2(b)]. In this case advection of the magnetic field in Eq.(18) is provided by the Hall term, while in Eq.(19) the quadrupole field generation by the Hall term is balanced by the plasma compressibility [rather than by the resistive diffusion as in the regime 2(a)]. Thus, Eq.(32) for the resulting magnitude of the quadrupole field $b$, and Eq.(42) for the reconnected flux $\psi_i$, should be now complemented with the relation (41) for the resistive term in Eq.(18), and a similar relation for the Hall term in this equation:

$$(\gamma \tau_A)^{-1} dkxb \sim \psi_i \tag{56}$$

Altogether, these equations yield:

$$(\gamma \tau_A) \sim \beta^{1/2} kd, (\Delta x) \sim \beta^{-1/4} S^{-1/2} (kd)^{-1/2}, \psi_i \sim \beta^{-1/4} S^{-1/2} d^{-1/2} k^{-3/2} \tag{57}$$

As seen from (57), in this case the instability growth rate, quite counterintuitively [and similarly to the regime (2a)], does not depend on the magnitude of the Lundquist number $S$. Note that these two non-constant-psi Hall-MHD regimes, 2(a) and 2(b), are entirely new theoretical findings.

The regime 2(b) should transfer into the single-fluid regime 1 when the Hall parameter $d$ falls below a certain threshold value $\tilde{d}_1$. The latter can be derived by estimating the magnetic field advection by the bulk flow of plasma [the first term on the r.h.s. of Eq.(18)] and comparing it with the advection caused by the Hall effect (the last term in this equation). Thus, it follows from Eqs. (20) and (22) that

$$\varphi'' \sim (\gamma \tau_A)^{-2} k (\Delta x) \psi'' \sim (\gamma \tau_A)^{-2} \Rightarrow \varphi \sim (\gamma \tau_A)^{-2} (\Delta x)^2 \tag{58}$$

Then, the bulk advection contribution can be estimated, with help of (57-58), as

$$kx\varphi \sim k(\Delta x)\varphi \sim k(\gamma \tau_A)^{-2} (\Delta x)^3 \sim S^{-3/2} \beta^{-7/4} k^{-5/2} d^{-7/2}, \tag{59}$$

which becomes significant when it equals the l.h.s. of Eq.(18), i.e. $\psi_i$. According to (57) and (59), it occurs at

$$d \sim \tilde{d}_1 \sim S^{-1/3} \beta^{-1/2} k^{-1/3} \tag{60}$$

Hence, at $d < \tilde{d}_1$, the instability enters the single-fluid regime 1 described by the scaling (43). On the other hand, in the between, i.e. at $\tilde{d}_1 < d < \tilde{d}_2$, there is also the transition from the regime 2(b) to the regime 2(a), which takes place at $\beta \sim 1$.

All five instability regimes, as well as their respective validity domains in the $(\beta - d)$ space, are shown in Fig.2. The non-constant-psi single-fluid regime 1 is described by the scaling (43). The non-constant-psi Hall-MHD regime 2(a) corresponds to the scaling (48). The EMHD constant-psi regime 3(a) has scaling given in Eq.(52). The intermediate constant-psi Hall-MHD regime 3(b) corresponds to the scaling (54). Finally, the intermediate non-constant-psi Hall-MHD regime 2(b) is described by the scaling (57).

For a very long-wave tearing mode with $k < S^{-1/3}$, and the Hall parameter $d \leq 1$ (which is of interest for a majority of applications) the number of possible instability regimes becomes reduced to three (see Fig.2). These are regimes 1 and 2(a,b). Note also that if $d > 1$, in deriving the external solution it is necessary to take into account the plasma inertia [17] [see condition (45)], which results in the reduction of the tearing parameter $\Delta'$.

## 4. Conclusion

Now one can get back to the issue raised in the Introduction, namely, what is the fastest growing tearing mode in the framework of the Hall-MHD theory. In general, it is quite a cumbersome task. Therefore, in what follows, we consider only the simplest example of the plasma with $\beta \geq 1$. As seen from Fig.1, in this case for the mode with a given wave-vector $k > k_*$ the Hall effect becomes significant if the Hall parameter $d > S^{-1/5} k^{1/5}$. Hence, the most susceptible to the Hall effect is the mode with the minimal $k$, i.e. the one with $k = k_* \sim S^{-1/4}$. Therefore, if

$$d < d_{min} \sim S^{-1/5} k_*^{1/5} \sim S^{-1/4}, \qquad (61)$$

all modes with $k > k_*$ remain in the single-fluid FKR regime 1. On the other hand, for modes with $k < k_*$ (see Fig.2), the threshold value of the Hall parameter is $d > S^{-1/3} k^{-1/3}$. Therefore, in this case the overall minimum of $d$ is defined by the mode with the maximal $k$, which is the very same mode with $k \sim k_*$. Thus, under the condition (61), all modes with $k > k_*$ are in the classical FKR regime 1 of Fig.1 governed by the scaling (26), while modes with $k < k_*$ are in the single-fluid Coppi regime 1 of Fig.2 described by the scaling (43). Therefore, since in the scaling (26) the instability growth rate is larger for smaller $k$, and vice versa is the case in the scaling (43), the mode with $k \sim k_*$ is the fastest one, and its increment $\gamma$ is equal to

$$\gamma_{max} \sim \tau_A^{-1} S^{-1/2} \qquad (62)$$

Consider now a situation when the Hall effect comes into play. According to (61), it requires $d > d_{min} \sim S^{-1/4}$, or

$$Z \equiv \frac{d}{d_{min}} = d S^{1/4} > 1 \qquad (63)$$

Then, as seen from Fig.1, the constant-psi Hall-MHD regime 2(a) takes the place within the interval $k_* < k < k_1$, where the upper boundary, $k_1$, is defined by the equation

$S^{-1/5}k_1^{1/5} \sim d$, i.e. $k_1 \sim Z^5 k_*$. Note that modes, occupying the interval $k_1 < k < 1$, remain in the single-fluid regime 1. Therefore, from all the modes within the domain $k_* \leq k < 1$, the fastest one is still the mode with $k \sim k_*$, which has now the growth rate

$$\gamma_{\max}^{(1)} \sim \tau_A^{-1} S^{-1/2} d^{1/2} k_*^{-1/2} \sim \tau_A^{-1} Z^{1/2} S^{-1/2} \tag{64}$$

Furthermore, in the domain $k < k_*$, which is illustrated in Fig.2, the modes with $k_2 < k < k_*$, where $k_2$ is defined by the condition $S^{-1}k_2^{-3} \sim d$, i.e. $k_2 \sim Z^{-1/3}k_*$, fall now into the Hall-MHD regime 3(a). Therefore, in this interval the maximal growth rate belongs to the mode with $k \sim k_2$, hence, according to (52),

$$\gamma_{\max}^{(2)} \sim \tau_A^{-1} S^{-1/2} d^{1/2} k_2^{-1/2} \sim \tau_A^{-1} Z^{2/3} S^{-1/2} \tag{65}$$

Further down the magnitude of the wave-vector $k$, the modes with $k_3 < k < k_2$, where $k_3$ is defined by the equation $S^{-1/3} k_3^{-1/3} \sim d$, which yields $k_3 \sim Z^{-3} k_*$, are in the regime 2(a). The very long-wave modes with $S^{-1} < k < k_3$ remain in the single-fluid regime 1. By comparing expressions (64) and (65), and recalling the scalings (43) and (48), respectively for the regimes 1 and 2(a), one concludes that in the Hall-MHD case the fastest mode has the wave-vector

$$k \sim k_2 \sim Z^{-1/3} S^{-1/4} \tag{66}$$

and the growth-rate given in Eq.(65).

Finally, as it was mentioned in Section 3, the diagram shown in Fig.2 holds only for the Hall parameter $d \leq 1$. This puts the upper limit on the magnitude of $Z$, namely $Z \leq Z_{\max} \sim S^{1/4}$. Thus, for $Z \sim Z_{\max}$, Eqs. (65-66) yield that for the fastest tearing mode

$$k \sim S^{-1/3}, \gamma \sim \tau_A^{-1} S^{-1/3} \tag{67}$$

This is a very long-wave mode. Therefore, in the context of the plasmoid instability, the issue is not only the fastest tearing mode per se, but also whether the aspect ratio of the current sheet under consideration is large enough to accommodate such a mode [1,3].

Clearly, all presented results and conclusions are relevant to situations when the Hall parameter $d \equiv d_i / L = c / \omega_{pi} L$ ($d \approx 3 \times 10^7 / L(cm)\sqrt{n_i(cm^{-3})}$ for a hydrogen plasma) is large enough [see Eq.(61)] to make the Hall effect involved in the reconnection process. Of course, the required inequality [which is opposite to the one in (61)] is satisfied in laboratory experiments dedicated to the Hall-MHD magnetic reconnection. Thus, in the MRX experiment[18], where typical parameters are equal to $n_i \sim 10^{14} cm^{-3}, L \sim 20 cm, T_e \sim 10 eV$ and $B \sim 10^2 G$, the Lundquist number is $S \sim 10^4$, while the Hall parameter $d \sim 0.1 - 0.3$.

Note, however, that in natural plasma objects this could be not the case even under a very large value of the Lundquist number. For example, in an active region of the solar corona, where $n_i \sim (10^9 - 10^{10}) cm^{-3}, T \sim 10^2 eV, B \sim 10^2 G$, L$\sim (10^9 - 10^{10}) cm$, one gets $S \sim 10^{12} - 10^{14}$, but only $d \sim 10^{-6}$, which is far too small because of a relatively high plasma density. Therefore, plasma objects with a lower number density of ions as, for example, the Earth's magnetotail, where $n_i \sim (10 - 10^2) cm^{-3}$, are more favourable environments for the Hall-mediated magnetic reconnection. However, for the very same reason, the collisionless model of reconnection, rather than the resistive one, seems to be more appropriate in such a case.

**ACKNOWLEDGMENTS**

This work was completed during the author's visit to the National Aviation Academy (NAA) in Baku. Thus, GV is grateful to R. Z. Sagdeev for the invitation to Baku, and to the whole staff at NAA for a very warm hospitality. A special thanks is also due to RZS for making helpful comments on the present paper.

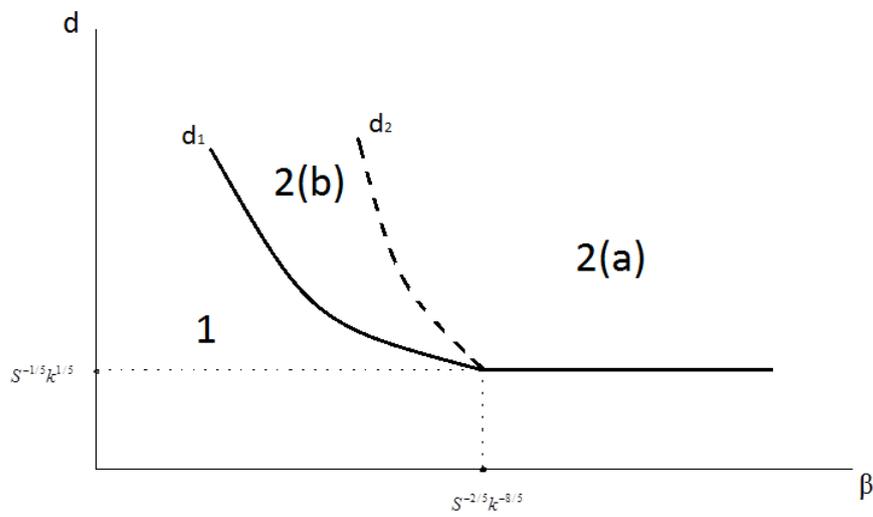

**Fig.1** Instability regimes for modes with $S^{-1/4} < k < 1$

    1 – the constant-psi single-fluid MHD regime [Eq.(26)]

    2 – the constant-psi Hall-MHD regimes

        (a)- the EMHD regime [Eq.(36)]

        (b)- the intermediate regime [Eq.(33)]

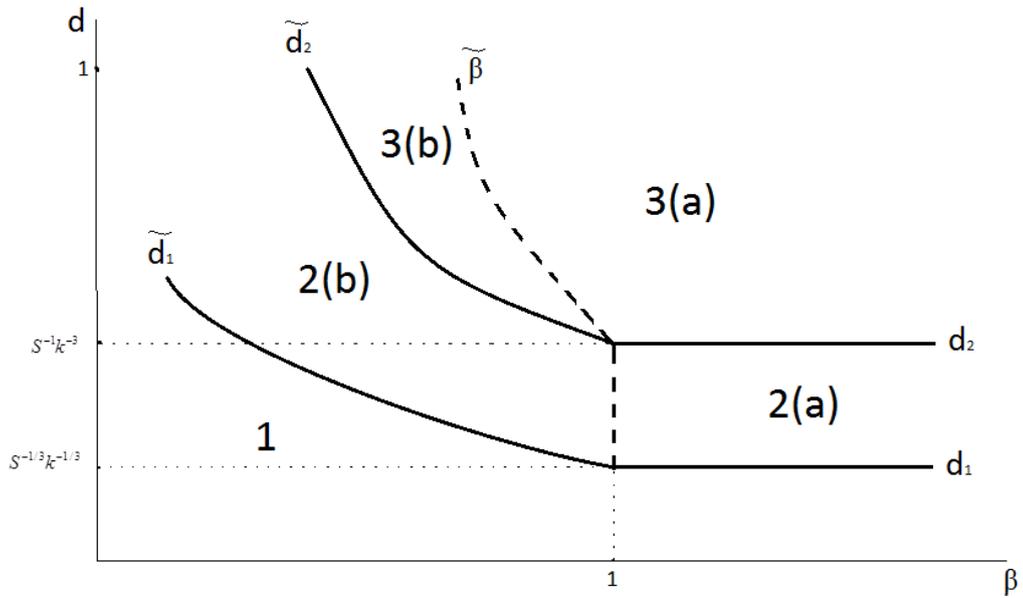

**Fig.2**  Instability regimes for modes with $S^{-1/3} < k < S^{-1/4}$

1 – the non-constant-psi single-fluid MHD regime [Eq.(43)]

2 – the non-constant-psi Hall-MHD regimes

  (a)- the EMHD regime [Eq.(48)]

  (b)- the intermediate regime [Eq.(57)]

3 – the constant-psi Hall-MHD regimes

  (a) – the EMHD regime [Eq.(52)]

  (b) – the intermediate regime [Eq.(54)]